\documentclass[twocolumn,english,aps,pra,showpacs]{revtex4}
\usepackage{amssymb}
\usepackage{amsmath}
\usepackage{graphicx}
\usepackage{epstopdf}
\usepackage{bm}
\usepackage{braket}
\usepackage{color}
\usepackage{rotating,booktabs}

\begin{document}
\title{Strongly Interacting Quantum Gases in One-Dimensional Traps}
\author{Li Yang$^1$, Liming Guan$^{2,1}$, and Han Pu$^{1,3}$}

\affiliation{$^{1}$Department of Physics and Astronomy, and Rice Quantum Institute,
Rice University, Houston, TX 77251, USA \\
$^2$Institute for Advanced Study, Tsinghua University,
Beijing, 100084,  P. R. China \\
$^3$Center for Cold Atom Physics, Chinese Academy of Sciences, Wuhan 430071, P. R. China}

\begin{abstract}
Under the second-order degenerate perturbation theory, we show that the physics of $N$ particles with arbitrary spin confined in a one dimensional trap in the strongly interacting regime can be described by super-exchange interaction. An effective spin-chain Hamiltonian (non-translational-invariant Sutherland model) can be constructed from this procedure. For spin-1/2 particles, this model reduces to the non-translational-invariant Heisenberg model, where a transition between Heisenberg anti-ferromagnetic (AFM) and ferromagnetic (FM) states is expected to occur when the interaction strength is tuned from the strongly repulsive to the strongly attractive limit. We show that the FM and the AFM states can be distinguished in two different methods: the first is based on their distinct response to a spin-dependent magnetic gradient, and the second is based on their distinct momentum distribution. We confirm the validity of the spin-chain model by comparison with results obtained from several unbiased techniques.
\end{abstract}

% 67.85.Lm: degenerate Fermi gases
% 75.10.Pq: spin chain models
% 75.30.Et: Exchange and superexchange interactions (see also 71.70.Gm Exchange interactions)
% 03.75.Mn: multicomponent and spinor condensates

\maketitle

\section{Introduction}
One dimensional (1D) quantum systems have received much attention during the past many decades. This is due to the fact that quantum effects are more pronounced in reduced dimensions, and also to the fact that many 1D models, such as the Lieb-Liniger model \cite{Lieb1963} and the Gaudin-Yang model \cite{Gaudin1967,Yang1967}, can be solved exactly with Bethe ansatz method \cite{Bethe1931,Guan2013}. Most exactly solvable models require the underlying systems to be translation invariant and the models can then be integrable. The presence of an external trapping potential in general breaks the integrability. One notable exception to this is a system of 1D spinless bosons with infinite contact repulsion (the so called Tonks-Girardeau gas) confined in an arbitrary trapping potential, which can be mapped into a non-interacting spinless Fermi gas \cite{Girardeau1960,Girardeau2001}, and has been realized in experiments using ultracold atoms \cite{Paredes2004,Kinoshita2004,Haller2009}. However, if the particles possess spin degrees of freedom, the problem becomes much more complicated. In recent years, there has been works on constructing the ground state of 1D spinful bosons and fermions with infinite or nearly infinite contact interaction \cite{Deuretzbacher2008,Guan2009,Girardeau2010,Girardeau2011,Fang2011}. It has been shown that, at exactly infinite interaction, the ground state of such spinful particles possesses degeneracy as the energy is independent of the spin configuration. Slightly away from this infinite repulsion limit, a perturbation theory can be constructed using $1/g$ (where $g$ is the strength of the contact interaction) as the small parameter. In this way, the ground state is governed by an effective  Hamiltonian defined within this degenerate subspace  \cite{Volosniev2014_1,Deuretzbacher2014}.

In this work, we will explicitly construct an effective model for 1D strongly interacting particles using a perturbation approach. Here the unperturbed system consists of particles with infinite contact interaction, i.e., $g=\infty$. At finite but large $|g|$, we take $1/g$ as the small perturbation parameter. We will show that we need to take the perturbation to second order in order to break all the spin degeneracy. In this way, we can construct an effective Hamiltonian which takes a form of a non-translational-invariant Sutherland model \cite{Sutherland1975}, which arises from the effective super-exchange interaction between neighboring particles. One can intuitively understand the emerence of the super-exchange term as follows. At $g=\infty$, particles are inpenetrable in 1D and they cannot exchange positions with their neighbors. Away from $g=\infty$, there will be small but finite probabiliy that two neighboring particles can exchange positions, which gives rise to the effective super-exchange interaction. For spin-1/2 fermions, which we focus on in this work, the exchange operator can be written in terms of spin operators, and the Sutherland model reduces to the Heisenberg model. It immediately follows that the ground state of spin-1/2 fermions is a Heisenberg anti-ferromagnetic (AFM) state in the strongly repulsive limit, and a ferromagnetic (FM) state in the strongly attractive limit (we exclude the tightly bound molecular states on the attractive side, i.e., we consider the upper branch of the system). We investigate the properties of such a system and demonstrate experimental signatures that allow us to distinguish the AFM and the FM states. using both the effective model and several unbiased methods, and show that the former is indeed valid in the strongly interacting regime.

The main advantages of the effective model are two fold. First, from a conceptual point of view, the effective model provides new insights to the quantum magnetic properties of strongly interacting particles in 1D. Second, from a practical point of view, the effective model is much easier to handle in comparison to unbiased methods. As a result, the effective model allows us to deal with more particle numbers and to investigate the dynamics to longer time scales. To this end, we benchmark our effective model against several unbiased methods and show that the former is indeed valid in the strongly interacting regime. These benchmark calculations also demonstrate that calculations based on the effective model are much more efficient and take much less time than those based on unbiased methods.

The rest of the paper is organized as follows. In Sec.~II, we derive the effective spin-chain Hamiltonian using a second order perturbation theory. We compare the energy spectrum obtained from this Hamiltonian with that obtained from a numerically exact Green's function method. In Sec.~III, we calculate the density profiles of the 1D trapped system in both real and momentum spaces. We show that the FM and the AFM states possess identical real space density profile, but with distinctive momentum distribution. In Sec.~IV, we study the system's response to a spin-dependent magnetic gradient, which breaks the SU(2) symmetry and hence mixes the AFM and the FM states. In Sec.~V, we show how the spin symmetry breaking term helps to realize the FM state in practice. Finally, in Sec.~VI, we discuss the advantages of the effective model over those unbiased methods, which serve as an important motivation for this work. Many of the technical details can be found in the Appendices.

\section{Effective spin-chain model}
We consider a one-dimensional system with $N$ strongly interacting spinful particles with mass $m$ trapped in an arbitrary external potential, with the Hamiltonian
\begin{equation}\label{originalHamiltonian}
H=\underbrace{\sum_{i=1}^N\left[-\frac{1}{2}\frac{\partial^2}{\partial x_i^2}+V(x_i)\right]}_{H_f}+\underbrace{g\sum_{i<j}\delta(x_i-x_j)}_{H_{\rm int}}\,.
\end{equation}
Here we have set $\hbar=m=1$. For infinite repulsion the particles become impenetrable and behave like spinless fermions. If the $N$ particles are spinless bosons, the many-body wave function can be constructed by Bose-Fermi mapping \cite{Girardeau1960}. For spinful fermions, the corresponding wave function can be generalized \cite{Deuretzbacher2008} to
\begin{equation}
\Psi(x_{1}\cdots x_{N},\sigma_{1}\cdots\sigma_{N})=\sum_{P}(-1)^{P}P\left[\varphi_{A}\theta^1\otimes\chi\right] \label{BFwave},
\end{equation}
where $\varphi_{A}$ is a Slater determinant which represents the eigen-wave function of $N$ spinless fermions governed by Hamniltonian $H_f$. Here $\theta^1$ is a sector function (i.e., generalized Heaviside step function) of spatial coordinates, whose value is one in spatial sector $x_1<x_2<\cdots<x_{N}$, and zero in any other spatial sectors. $\chi$ is a spin wave function, and $P$ is the permutation operator whose convention of acting on spatial and spin wave functions is presented in Appendix~\ref{con}.
%One thing to note is that for Heisenberg models derived from Hubbard model in Mott insulator phase \cite{Auerbach1998}, the wavefunction showing magnetism is simply to replace $\varphi_A\theta^1$ in Eq.(\ref{BFwave}) by $(1/\sqrt{N!})\prod_{j}w_{j}(x_{j})$, where $w_{j}(x)$ is the wannier function for site $j$.

To obtain an effective Hamiltonian for spinful fermions in the strongly interacting regime, we use the perturbation theory. To this end, we consider $H_f$ as the perturbation, and $H_{\rm int}$ as unperturbed Hamiltonian. This is in the same spirit as the procedure for constructing the effective spin model from the Hubbard model in the large interaction limit \cite{Auerbach1998}. The unperturbed Hamiltonian $H_{\rm int}$ has a degenerate ground state subspace with zero eigen-energy $E_{\rm int}^{(0)}=0$. This subspace is the space of all the anti-symmetric wave functions satisfying the boundary condition $\Psi{}_{x_{i}=x_{j}}=0$ \cite{Girardeau1960,Deuretzbacher2008}. Equation~(\ref{BFwave}) with a full set of $\varphi_A$'s constitute a complete basis for this subspace. We define a projection operator ${\cal P}_0$ into this subspace and its complementary operator ${\cal P}_1=1-{\cal P}_0$. Now let us consider the effect of $H_f$ on this subspace under the framework of degenerate perturbation theory. The first order effective Hamiltonian reads $H^{(1)}={\cal P}_0H_f{\cal P}_0$. The ground states of $H^{(1)}$ still form a degenerate subspace whose eigen-vectors take the same form as Eq.~(\ref{BFwave}) with $\varphi_A$ representing the lowest-energy Slater determinant for $H_f$. (From now on, we denote $\varphi_A$ as such a lowest-energy Slater determinant.) To lift the remaining spin degeneracy, we therefore have to carry out the perturbation calculation to second order. Let ${\cal Q}_0$ be the projection operator into the ground state subspace of $H^{(1)}$. Applying standard degenerate perturbation theory, we obtain the second-order effective Hamiltonian as (see Appendix~\ref{deg} for details).
  \begin{equation}
  H^{(2)}={\cal Q}_0H_f{\cal P}_1\frac{1}{E_{\rm int}^{(0)}-H_{\rm int}}{\cal P}_1H_f{\cal Q}_0.
  \end{equation}
  After some algebra (for details, see Appendix~\ref{spin}), we find that, after neglecting a constant ${\cal Q}_0H_f{\cal Q}_0$, the effective second-order Hamiltonian can be written as
  \begin{equation}
  H_{\rm eff}=-\frac{1}{g}\sum_{i=1}^{N-1}C_i (1-{\cal E}_{i,i+1}) \label{effectiveH},
  \end{equation}
where ${\cal E}_{i,i+1}$ is the exchange operator acting on a spin state $\chi$ within the subspace defined by ${\cal Q}_0$, and its effect is to exchange the $i^{\rm th}$ and $(i+1)^{\rm th}$ particles, and the coefficients
\begin{equation}
\label{ci} C_i=N!\int\prod_{j}dx_{j}\left|\partial_{i}\varphi_{A}\right|^{2}\delta(x_{i+1}-x_{i})\theta_{[i+1,i]}^{1} \,,
\end{equation}
are positive constants independent of spin, where \[ \theta_{\left[i,i-1\right]}^{1}=\theta^1/\theta(x_i-x_{i-1})\,,\] is a reduced sector function (see Appendix~\ref{spin}). $H_{\rm eff}$ takes the form of the non-translational-invariant Sutherland model, and physically arises from the effective super-exchange interaction when $g$ deviates away from infinity, as we have mentioned earlier. In the case of spinful bosons, following the same procedure leads an effective Hamiltonian similar to (\ref{effectiveH}) with the minus sign before ${\cal E}_{i,i+1}$ replaced by the plus sign. This spin-chain model preserves the SU(2$s$+1) symmetry, where a single particle has spin $s$. Being a bipartite Hamiltonian, the Lieb-Mattis theory \cite{Lieb1963,Auerbach1998} is also satisfied. Since it is made up of permutation operators, it can also be block diagonalized in the irreducible representation of the permutation group $S_{N}$ \cite{Mila2014,Ma2007}.

We comment here that Eq.~(\ref{BFwave}) can be written in a different form, $\Psi=\varphi_{A}\sum_{P}c_{\{\sigma\},P}P\theta^1$, with $c_{\{\sigma\},P}=1/(N_{\uparrow}!N_{\downarrow}!) \braket{\{\sigma\}|P\chi}$ being weights in different sectors for a spin configuration $\{\sigma\}$. These weights can be regarded as variational parameters and determined by $\partial E/\partial c_P=0$ together with the Bethe-Peierls boundary condition \cite{Volosniev2014_1,Volosniev2014_2}. For strong but finite interaction, the eigen-energies read $E=E_0-K/g+O(1/g^2)$, here $K$ is the Tan contact \cite{Tan2008}. An effective spin model can be constructed from this variational approach, as done by several groups \cite{Volosniev2014_1,Volosniev2014_2,Deuretzbacher2014,Levinsen2014}. Our result based on the perturbation calculation is consistent with these results.

\begin{figure}[h]
\includegraphics[width=8.cm]{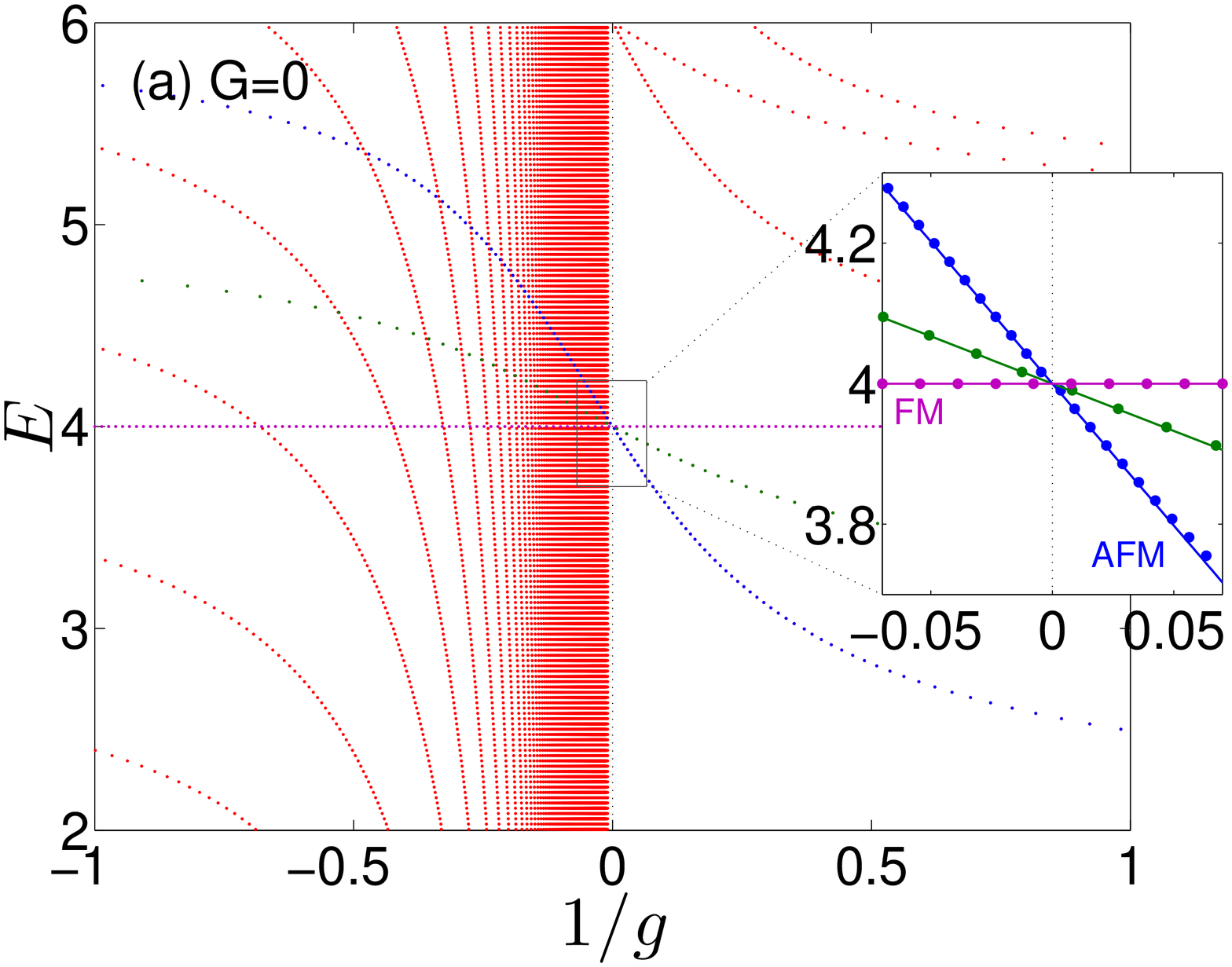}
\includegraphics[width=8.cm]{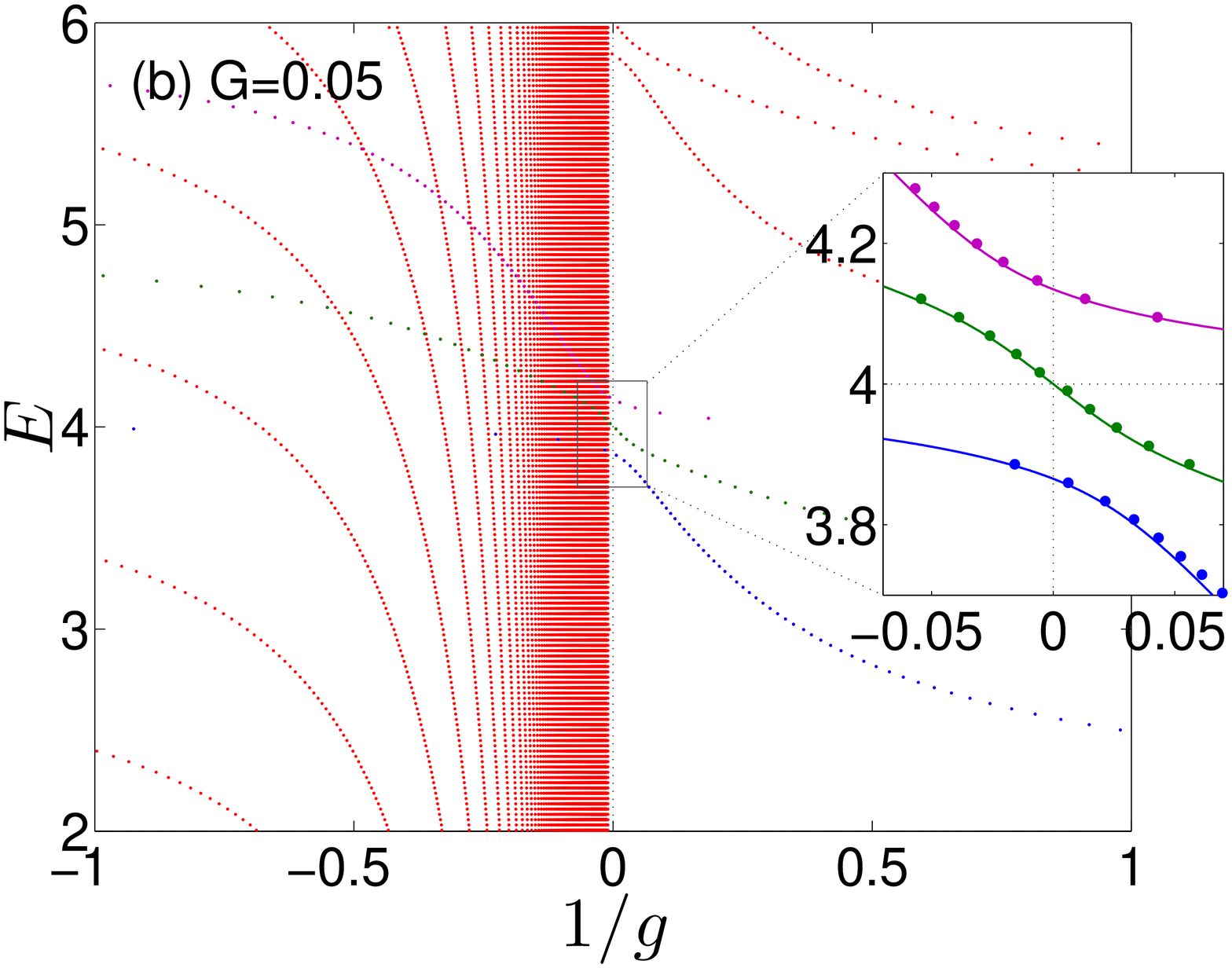}
\caption{(Color online) Energy spectrum of the relative motion as a function of $1/g$ for three fermions with $(N_{\uparrow}, N_{\downarrow})=(1,2)$, without (a) and with (b) the spin-dependent magnetic gradient. For (b), we have $G=0.05$. The main figures are obtained using the Green's function method. The red dotted lines in negative $g$ area represent the tightly bound molecular states. The inset figures show the comparison between the spectrum obtained from the Green's function method (dots) and that from the effective spin-chain model (solid lines) near $1/g=0$. In all the figures presented in this paper, we have adopted the trap units with $\hbar=m=\omega=1$. Consequently, the energy $E$ is in units of $\hbar \omega$, and the interaction strength $g$ is in units of $\sqrt{\hbar^3 \omega/m}$.}\label{Eg1N3}
\end{figure}

To benchmark the spin-chain model, we show in Fig.~\ref{Eg1N3}(a) the low energy spectrum of a three-body system. Similar benchmarks were also performed in Refs.~\cite{Deuretzbacher2014,Levinsen2014}. In this work, we focus on spin-1/2 fermions, and label the two spin species as $\uparrow$ and $\downarrow$. The external potential is chosen to be a harmonic potential with frequency $\omega$. In our calculation, we take $\omega=1$ along with $\hbar$ and $m$, and the observables are normalized to dimensionless values: $x\sim x/\sqrt{\hbar/(m\omega)}$, $p\sim p/\sqrt{\hbar m\omega}$, and $E\sim E/\hbar\omega$. The main figure of Fig.~\ref{Eg1N3}(a) is obtained by the unbiased Green's function method based on the original many-body Hamiltonian (\ref{originalHamiltonian}) \cite{Blume2013}. In the inset, we compare this exact spectrum (dots) with the spectrum obtained from the spin-chain Hamiltonian $H_{\rm eff}$ (solid lines). As one can see, in the strong interaction regime with $1/ |g| \ll 1$, the spin-chain model faithfully reproduces the exact spectrum of the upper branch when the tightly bound molecular states on the attractive ($g<0$) side are ignored.

We can gain some insights into the spectrum of $H_{\rm eff}$ by noting that the eigenvalues of the exchange operator ${\cal E}_{i,i+1}$ are $\pm 1$. Therefore, for $g>0$, the spectrum of $H_{\rm eff}$ has a lower bound of $-(2/g)\sum_{i=1}^{N-1}C_i$ (corresponding to a fully anti-symmetric spin configuration with ${\cal E}_{i,i+1}=-1$ for any $i$), and an upper bound of 0 (corresponding to a fully symmetric spin configuration with ${\cal E}_{i,i+1}=1$ for any $i$). We remark that the fully anti-symmetric spin configuration can only be realized for $2s+1 \ge N$. For not too small $N$, this requires a fermionic species with large spin $s$. Recent cold atom experiments have witnessed realization of high spin Fermi gases in alkali-earth atoms \cite{high1,high2,high3,high4}. For $g<0$, the spectrum is inverted and bound between 0 and $|2/g|\sum_{i=1}^{N-1}C_i$.

\section{Density Profiles in Real and Momentum Spaces}
Let us now examine in detail the density profiles in both real and momentum spaces for the ground state of $H_{\rm eff}$. For spin-1/2 fermions, the exchange operator can be written in terms of the spin operators: \[{\cal E}_{i,j}=(1+\vec{\sigma}_i\cdot \vec{\sigma}_j)/2\,,\] where $\vec{\sigma}_i$ are the Pauli spin matrices for the $i$th atom. Hence we can rewrite the effective Hamiltonian (\ref{effectiveH}) as 
\begin{equation} 
H_{\rm eff}= -\frac{1}{g} \sum_{i=1}^{N-1} \,C_i (1-\vec{\sigma}_i\cdot \vec{\sigma}_{i+1})/2 \,,\label{heff1} \end{equation} 
which takes the form of the non-translational-invariant Heisenberg model with $C_i/(2g)$ plays the role of the super exchange coefficient between the $i$th and the $(i+1)$th spin. The effective spin-spin interaction is ferromagnetic for $g<0$, and anti-ferromagnetic for $g>0$. We therefore label the corresponding ground state FM for $g<0$ and AFM for $g>0$, as shown in the inset of Fig.~\ref{Eg1N3}(a), which is consistent with the Bethe ansatz result for the homogeneous case \cite{Guan2007,Oelkers2006}. Note that as the number of atoms in each spin species are individually conserved, the spin configuration for the FM state here can be written as $(S^{-})^{N_{\downarrow}}\ket{\uparrow\uparrow\cdots\uparrow}$, with $S^-=\sum_i \sigma^-_i/2$ being the total spin lowering operator.

To find the density profiles in both real and momentum spaces, let us first introduce the one-body density matrix element defined as
\begin{widetext}
\begin{equation}
\rho_{\sigma'\sigma}(x',x)
=\sum_{\sigma_2\cdots\sigma_N}\int dx_2\cdots dx_N \, \Psi^*(x',x_2\cdots x_N,\sigma',\sigma_2\cdots\sigma_N)\Psi(x,x_2\ldots x_N,\sigma,\sigma_2\cdots\sigma_N) \,,
\end{equation}
\end{widetext}
from which the real-space and momentum space density profiles can be calculated as
\begin{eqnarray*}
\rho_\sigma(x) &=& N \rho_{\sigma, \sigma}(x,x) \,,\\
\rho_\sigma(p) &=& (N/2\pi)  \int dx \int dx'\,e^{-ip(x-x')}\rho_{\sigma,\sigma}(x',x)\,.
\end{eqnarray*}
In Appendix~\ref{den}, we provide the details of calculating the one-body density matrix element given a many-body wave function as in Eq.~(\ref{BFwave}).

\begin{figure}[h!]
\includegraphics[width=8.5cm]{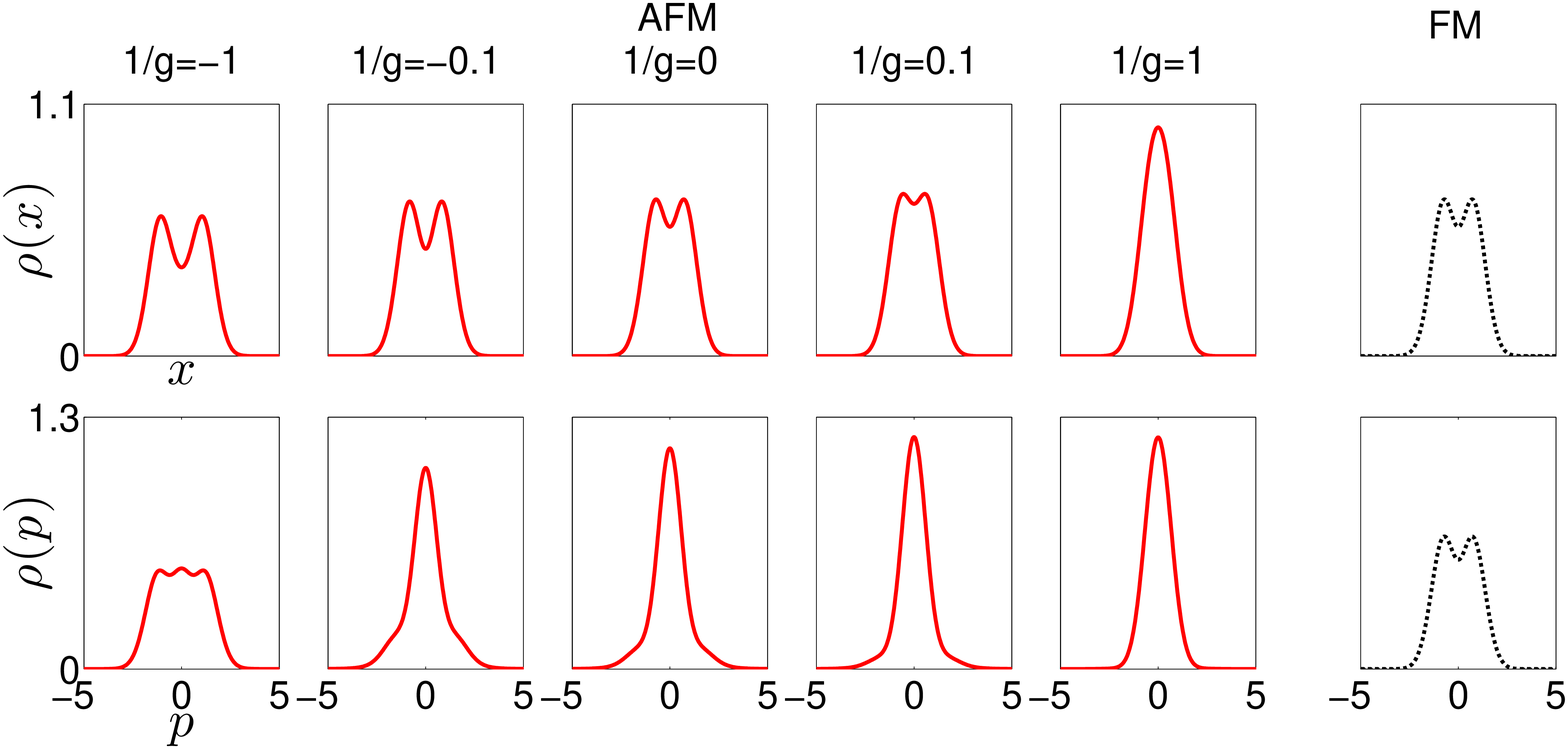}
\caption{(Color online) Real space density profiles (upper panel) and momentum space density profiles (lower panel) for $(N_{\uparrow}, N_{\downarrow})=(1,1)$ versus $1/g$ for AFM (red solid lines) and FM (black dotted lines) states. In our trap units, position $x$ is units of $\sqrt{\hbar/(m\omega)}$, the real space density $\rho(x)$ is in units of $\sqrt{m\omega/\hbar}$, the momentum $p$ is in units of $\sqrt{\hbar m \omega}$, and the momentum space density $\rho(p)$ is in units of $1/\sqrt{\hbar m \omega}$.}\label{mom_profile_N2}
\end{figure}

In Fig.~\ref{mom_profile_N2}, we present the density profiles for $N=2$ spin-1/2 fermions with $(N_{\uparrow}, N_{\downarrow})=(1,1)$. For this two-body problem, exact analytic solutions for arbitrary interacton strength $g$ can be found \cite{Busch1998}. Results desplayed in Fig.~\ref{mom_profile_N2} are obtained from the exact method. As a result, we are not limited to large $|g|$. Note that the FM state corresponds to a fully symmetric spin configuration $\chi$, and its density profiles, which are $g$-independent, are identical to a system of $N$ spinless fermions. More specifically, $\rho_\sigma(x)=(N_\sigma/N)\sum_{i=0}^{N-1}|\phi_i(x)|^2$, where $\phi_i(x)$ is the $i$th eigen-wave function of the single particle Hamiltonian; and $\rho_\sigma(p)$ decays as $\exp(-p^2)$ in the large $p$ limit.

The AFM state, on the other hand, possesses a fully anti-symmetric spin configuration and its density profiles are sensitive to the value of $g$. As $1/g \rightarrow 0$, the real-space density profile of the AFM state approaches that of the FM state, whereas the momentum space density profile remains distinct for these two states. Hence, in the strongly interaction limit, the density profiles for the AFM and the FM states are indistinguishable in real space, but distinguishable in momentum space. This statement remains true for $N>2$.

\begin{figure}[h!]
\includegraphics[width=8.5cm]{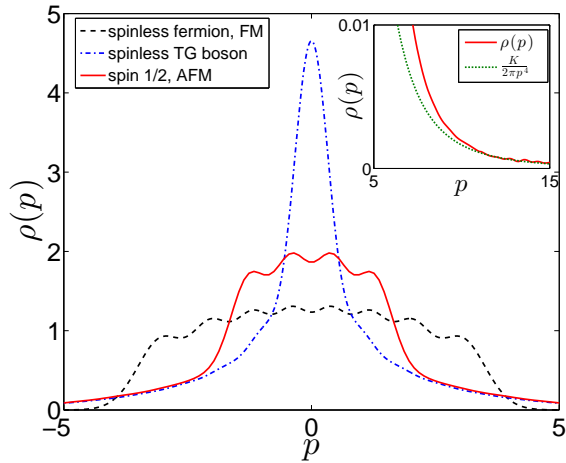}
\caption{(Color online) Momentum distribution for $(N_{\uparrow},N_{\downarrow})=(4,4)$. The black dashed curve is for the fully spin symmetric FM state, which has the same momentum distribution as $N$ spinless fermions. The red solid curve is for the AFM state. The blue dash-dotted curve is for the fully spin anti-symmetric state, which has the same momentum distribution as $N$ spinless Tonks-Girardeau bosons. The inset shows the momentum distribution for the AFM state in the large momentum limit, in comparison to the theoretical prediction $K/(2\pi p^4)$ (green dotted curve), where $K$ is Tan contact. }\label{mom_profile_N8}
\end{figure}

As a further example, we consider a system of $(N_{\uparrow},N_{\downarrow})=(4,4)$ spin-1/2 fermions in the strongly interacting limit. In Fig.~\ref{mom_profile_N8} we show the momentum space density profiles. The black dashed line corresponds to the momentum distribution of the FM state (which is the same as the momentum distribution of $N$ spinless fermions), and the red solid line to that of the AFM state. The AFM state has a nonzero Tan contact $K$, and in the large momentum limit, we have $\rho(p) = K/(2\pi p^4)$ \cite{Barth2011}. This is confirmed by our numerics as shown in the inset of Fig.~\ref{mom_profile_N8}. For comparison, we also show the momentum distribution of a fully anti-symmetric spin state, which coincides with the momentum distribution of $N$ spinless bosons in the Tonks-Girardeau limit. As we mentioned earlier, the fully anti-symmetric spin state is only possible when $2s+1 \ge N$ \cite{Yang2011}. We emphasize again that these different states have identical real space density profile, but can be distinguished from their distinctive momentum distribution.

\section{RESPONSE TO SPIN-DEPENDENT MAGNETIC GRADIENT}
The form of the spin-chain effective Hamiltonian $H_{\rm eff}$ makes it clear that a quantum phase transition is induced as $1/g$ is tuned across zero, which can be achieved using the technique of confinement induced resonance \cite{cir1,cir2}. In practice, however, more effort is required to observe this phase transition. The AFM ground state for $g>0$ can be straightforwardly prepared. Such is not the case for the FM state on the attractive side with $g<0$. This is due to the fact that, for $g<0$, there exist many bound molecular states with lower energies than the FM state, as can be seen from Fig.~\ref{Eg1N3}(a). If one simply prepare the system on the attractive side, these molecular states, not the FM state, will be realized. Hence to create the FM state, one needs to start from the AFM state on the repulsive side and adiabatically tune the interaction strength to the attractive side. However, the spin states are protected by symmetry: If we start from the AFM state and tune $1/g$ across zero, the system will remain as an AFM state and realize a fermionic super-Tonks-Girardeau state \cite{Guan2010,Astrakharchik2005}, as there is no coupling between the AFM and the FM states. To overcome this problem, we need to add a spin symmetry breaking term. One possibility is to add a spin-dependent gradient term. We will consider in detail how to realize the FM state in the next section. Here we first investigate how the AFM and the FM states respond to such a gradient term.

To this end, we introduce a weak spin-dependent magnetic gradient which adds a term $-G\sum_i x_i\sigma_i^z$ to the Hamiltonian (\ref{originalHamiltonian}), where $G$, which we will take to be non-negative, chracterizes the magnitude of the magnetic gradient. The effective spin-chain Hamiltonian will be modified corresondingly as
\begin{equation}
H_{\rm eff}=-\frac{1}{g}\sum_{i=1}^{N-1}C_i (1-\vec{\sigma}_i\cdot \vec{\sigma}_{i+1})/2  -G\sum_{i=1}^N D_i\sigma_i^z \,,
\label{heff}
\end{equation}
where $D_i=N!\int  x_i |\varphi_A|^2\theta^{1}\prod_{j=1}^N dx_j$ represents the position of the $i$th atom. In Fig.~\ref{Eg1N3}(b), we plot the energy spectrum for a three particle system in the presence of weak spin gradient, obtained from both the Green's function method and the effective model. Again we see excellent agreement in the strongly interacting regime. Comparing the insets of Fig.~\ref{Eg1N3}(a) and (b), one can easily see that the gradient term lifts the spin degeneracy at $1/g=0$, and the ground state is now separated from excited states by a finite gap, which facilitates the adiabatic preparation of the FM state to be discussed later.

The spin gradient tends to separate the two spin species \cite{Cui2014}. To quantify this effect, we define
\begin{equation}
\Delta=\frac{1}{N}\, \sum_{i=1}^N \langle x_i\sigma_i^z\rangle \,,
\end{equation}
which measures the center-of-mass separation between the two spin species. Here the expectation value is taken with respect to the ground state of the effective Hamiltonian (\ref{heff}). In the absence of the gradient ($G=0$), $\Delta=0$ for both the FM and the AFM states. Under the effective spin-chain model, $\Delta$ is a function of $Gg$ only.

\begin{figure}[h!]
\includegraphics[width=8.5cm]{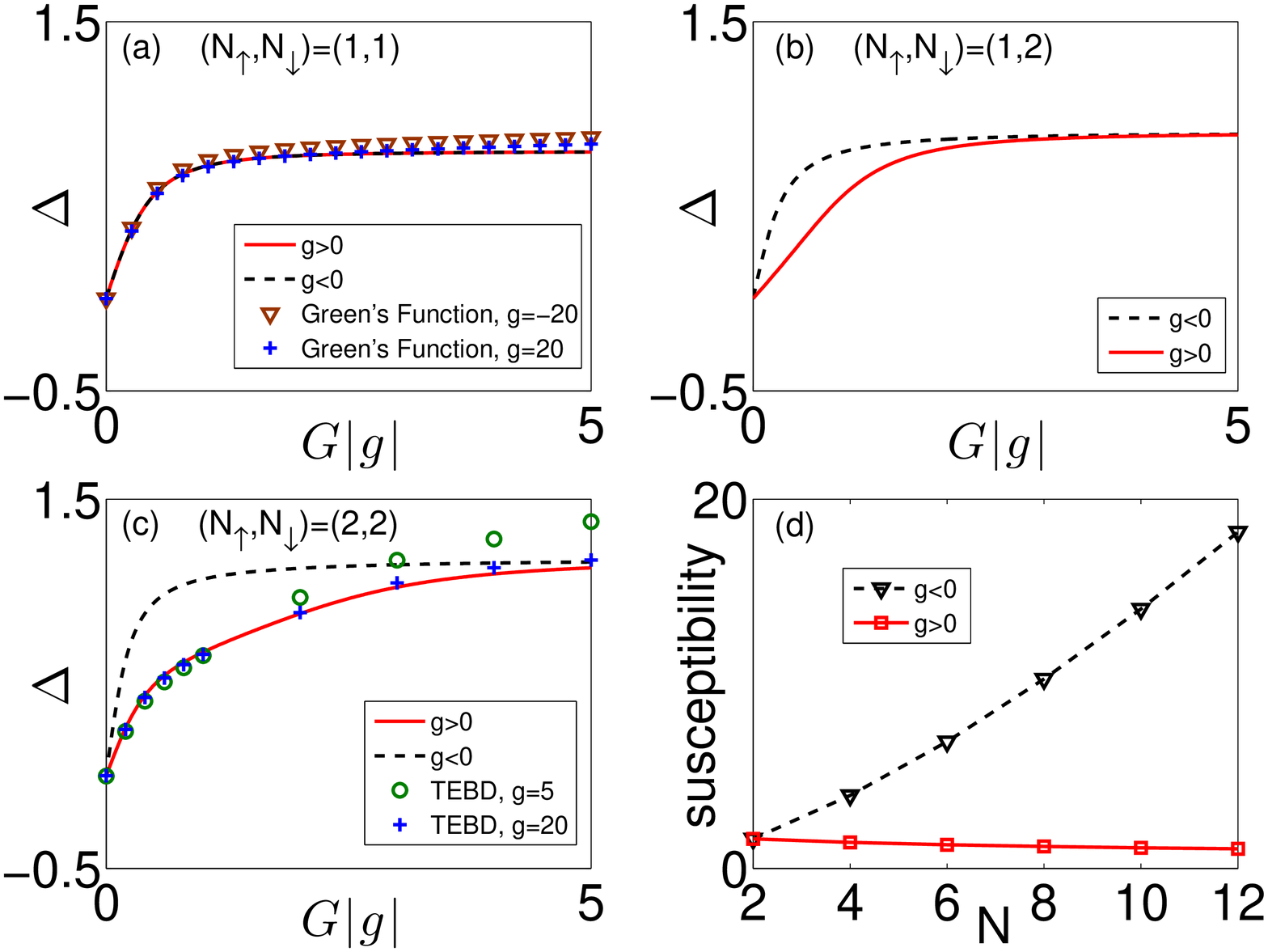}
\caption{(Color online) Separation between the two spin species as a function of $G|g|$ for (a) $(N_{\uparrow},N_{\downarrow})=(1,1)$, (b) $(N_{\uparrow},N_{\downarrow})=(1,2)$, (c) $(N_{\uparrow},N_{\downarrow})=(2,2)$. The black dashed curves are for the ground state with negative $g$, and the red solid curves are for the ground state with positive $g$. The symbols in (a) are obtained from the analytic solution detailed in Appendix~\ref{2b}. The symbols in (c) are TEBD results. (d) The susceptibility $\frac{d \Delta}{|g|d G}\Big|_{G=0}$ as functions of $N$ for $N_{\uparrow}=N_{\downarrow}=N/2$. In our trap units, $\Delta$ is in units of $\sqrt{\hbar/(m\omega)}$, and $Gg$ in units of $\hbar^2\omega^2$.}\label{deltaGgfig}
\end{figure}

As a first example, we again consider a two particle system with $(N_\uparrow, N_\downarrow)=(1,1)$. For this simple system, Hamiltonian (\ref{heff}) can be easily diagonalized, and $\Delta$ has an analytic expression:
\[ \Delta=\sqrt{\frac{2}{\pi}}\frac{\left[2G|g|+\sqrt{1+4(G|g|)^{2}}\right]^{2}-1}
{\left[2G|g|+\sqrt{1+4(G|g|)^{2}}\right]^{2}+1} \,.\]
Note that since $\Delta$ only depends on $|g|$, we conclude that the FM and the AFM state respond identically to the gradient in the two-body case. We plot $\Delta$ as a function of $G|g|$ in Fig.~\ref{deltaGgfig}(a). In the figure, we also plot the result obtained from an exact solution using the Green's function method with $g=\pm 20$, which are in good agreement with the effective model. The details of this solution can be found in Appendix~\ref{2b}.

By contrast, for $N>2$, the ground states for $g>0$ and $g<0$ will response differently to the gradient. In Fig.~\ref{deltaGgfig}(b) and (c), we plot $\Delta$ as a function of $G|g|$ for the cases $(N_\uparrow, N_\downarrow)=(1,2)$ and (2,2), respectively. The dashed and solid curves correspond to the ground state of negative and positive $g$, respectively. In general, the ground state on the attractive side will have a stronger response. To benchmark the effective model, we studied this problem using the Time-Evolving Block Decimation (TEBD) method \cite{Schollwck2011,Vidal2003,Tezuka2010,Wall2009}. In TEBD, a many-body wave function is represented by a Matrix-Product state (MPS), which approximates a many-body wave function by making a truncation of the entanglement spectrum. For 1D gapped system, whose entanglement is short-ranged, the truncation error is well controlled, and the TEBD method therefore represents an unbiased method and has been implemented widely to study 1D systems. The symbols in Fig.~\ref{deltaGgfig}(c) are the TEBD results for positive $g$. One can see that for large $g$, the results obtained from TEBD and the effective model agree with each other very well.

To further quanitify the response to the gradient and show the difference between the AFM and the FM states, we define the magnetic gradient susceptibility as $\frac{1}{|g|}(d\Delta/dG)_{G=0}$, and the following relation can be readily derived:
\begin{equation}
\frac{1}{|g|}\frac{d \Delta}{d G}\Big|_{G=0}=\frac{2}{|g|N^2}\,\sum_{n\neq 0}\frac{\left|\langle 0 |\sum_{i=1}^{N}x_i\sigma_i^z|n\rangle \right|^2}{E_n-E_0}\,,
\end{equation}
where $|n\rangle$ represents the $n$th eigenstate of the spin-chain Hamiltonian with $G=0$, and $E_n$ is the corresponding eigen-energy. $|0 \rangle$ represents the ground state, which is the AFM (FM) state for positive (negative) $g$. In Fig.~\ref{deltaGgfig}(d) we plot this susceptibility as a function of the total particle number $N$ for the case with $N_\uparrow=N_\downarrow=N/2$. One can see that, as long as $N>2$, the FM state possesses a larger susceptibility, i.e., is more prone to spin segregation under the gradient, than the AFM state. Furthermore, the susceptibility for the FM state grows rather rapidly as $N$ increases, whereas that for the AFM state is not very sensitive to $N$.

\section{Adiabatic preparation of ferromagnetic state}

\begin{figure}[h!]
\includegraphics[width=8.5cm]{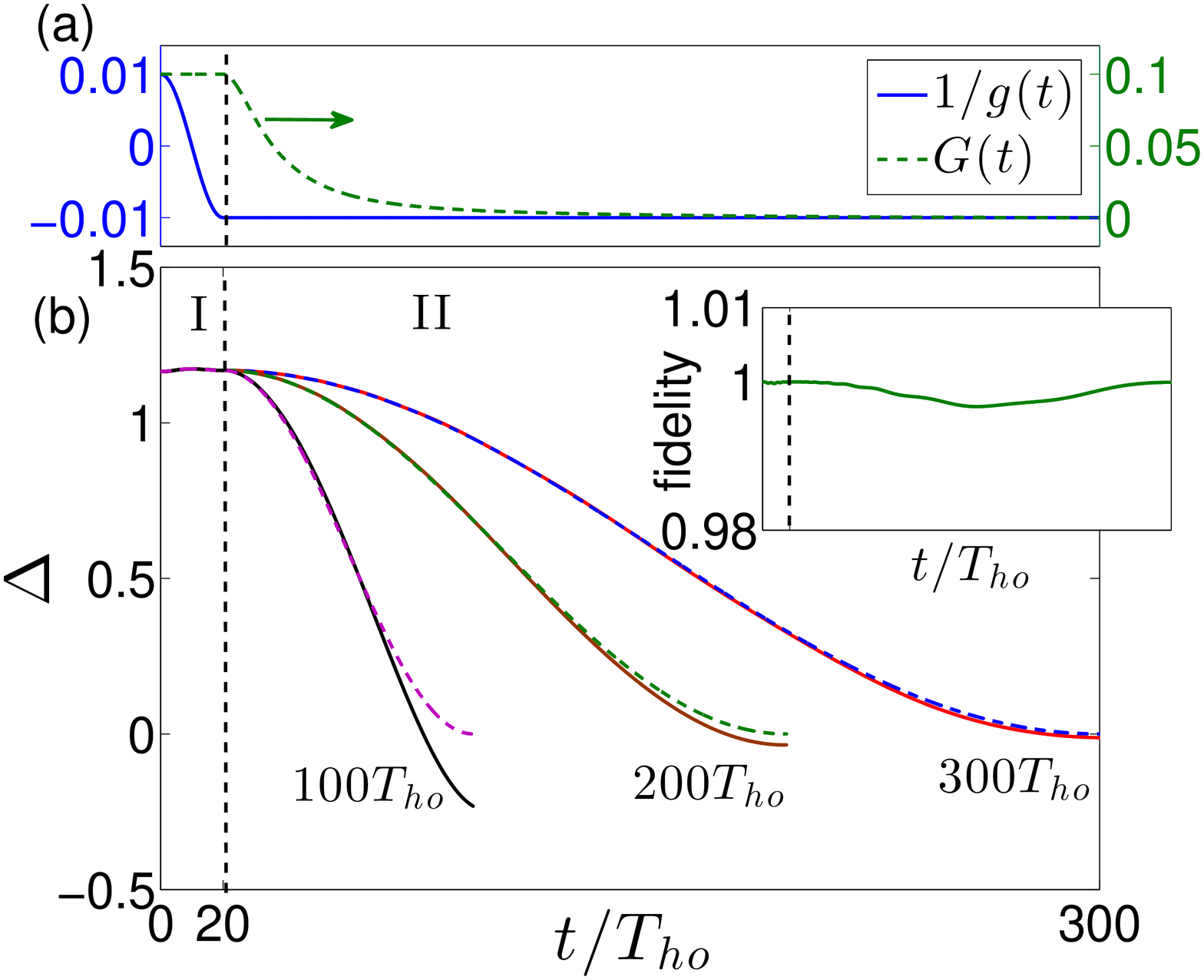}
\caption{(Color online) Adiabatic preparation of the FM state. At $t=0$, the system is prepared in the ground state with $1/g=0.01$ and $G=0.1$. (a) The value of experimentally controlled parameters $1/g(t)$ and $G(t)$ for a total adiabatic evolution time $300T_{\rm ho}$. (b) The solid lines represent $\Delta(t)$ obtained by solving the time-dependent Schr\"{o}dinger equation under the effective Hamiltonian $H_{\rm eff}$. The dashed lines represent Eq.~(\ref{del}), which is the $\Delta(t)$ of the instantaneous ground state for the given values of $g(t)$ and $G(t)$. Three different total adiabatic evolution time is calculated, $100\,T_{\rm ho}$,$200\,T_{\rm ho}$, and $300\,T_{\rm ho}$. The inset shows the fidelity of the adiabatically prepared state for the total evolution time $300\,T_{\rm ho}$.}\label{adiabatic_preparation}
\end{figure}

In the previous section, we suggested a method of applying weak spin-dependent magnetic gradient to approach the FM state in experiment. Here we will discuss the method in detail. The experimental protocol is in the following: (1) The system is initially prepared in the ground state with strong repulsion $(g>0)$ and a relatively large magnetic gradient. In the example presented in Fig.~\ref{adiabatic_preparation}, we choose the initial values $1/g=0.01$ and $G=0.1$. (2) From $t=0$ to $T_1$, $G$ is fixed at the initial value while the interaction strength is tuned to the attractive side as $1/g(t)=0.01\cos({\pi  t}/{T_1})$, which can be achieved with  confinement-induced-resonance method. (3) Finally, from $t=T_1$ to $T_1+T_2$, $g$ is fixed at its value at $T_1$, while the gradient strength $G$ is slowly turned off. We vary $G$ such that the instantaneous spin separation $\Delta$ follows the form \begin{equation}
\Delta(t)=\Delta[G(t)]=\Delta(T_1) \,\cos^2 \left[\frac{\pi (t-T_1)}{2T_2}\right] \,.
\label{del}
\end{equation}
The experimentally controlled parameters are plotted in Fig.~\ref{adiabatic_preparation}(a) for $T_1=20\,T_{\rm ho}$ and $T_2=280\,T_{\rm ho}$, where $T_{\rm ho}=2\pi/\omega$ is the harmonic trap period.

In Fig.~\ref{adiabatic_preparation}(b) we display the evolution of the spin separation parameter $\Delta$ in an example system with $(N_{\uparrow},N_{\downarrow})=(2,2)$, $T_1=20\,T_{\rm ho}$, and $T_1+T_2=100\,T_{ho}$, $200\,T_{ho}$, and $300\,T_{ho}$. The dashed curves represent the targetted instantaneous value as shown in Eq.~(\ref{del}); while the solid curves are obtained by solving the time-dependent Schr\"{o}dinger equation under the effective Hamiltonian $H_{\rm eff}$. As expected, the larger the $T_2$, the better agreement between the solid and dashed curves. In the inset, we also show the fidelity, which is the overlap between the calculated wave function from evolving the Schr\"{o}dinger equation and the instantaneous ground state wave function given the values of $g$ and $G$ at the moment, for the case $T_1+T_2=300\,T_{\rm ho}$. One can see that an FM state can be realized with very high fidelity. For a shorter total evolution time with $T_1+T_2=100\,T_{\rm ho}$, we still obtain a fidelity higher than 94\%.

Although we have proposed to use a spin-depedent magnetic gradient to break the spin symmetry and facilitate the adiabatic preparation of the FM state, in reality any spin symmetry breaking term can do the job. Experimentally, this mean one needs to introduce some perturbation to the system to which the two atomic spin states will respond differently. A possibility is to apply an off-resonant light with proper polarization such that it induces different light shift to different atomic spin states. This idea has been recently implemented to create spin-dependent optical lattices for cold atoms \cite{spin1,spin2}. 

Finally, we comment on the stability of the FM state. Due to presence of the tightly bound molecular states on the attractive side, the FM can only be metastable. In 2009, Haller {\em et al.} realized such a metastable state in a system of spinless bosons \cite{Haller2009}, and the resulting state is the so called super Tonks-Girardeau (sTG) gas. In that experiment, a typical lifetime of about 100 ms is found. We expect the lifetime of the FM state in a spin-1/2 Fermi gas should be longer than the bosonic sTG gas. This is because the low-lying molecular states for fermions must be spin singlet. Therefore the spin symmetric FM state will be protected by spin symmetry against decaying into the molecular states.

%In experiment the typical trapping frequency is about $\omega=2\pi \times 1 kHz$,correspondingly $300T_{ho}=300ms$. The lifetime of metastable FM states can be estimated from the lifetime of the metastable super Tonks-Girardeau(sTG) gases as they have the similar energy spectrum at large $|g|$. In 2009 people have realized the sTG state in spinless cesium gases and observed a typical lifetime of about 100ms of sTG states \cite{Haller2009}. However, in the region that the spin chain model approximation is good, the adiabatic time for the same performance as our calculation of Fig.\ref{adiabatic_preparation} is proportional to $(m^2a_{ho}^3g)/\hbar^3$. So for example if half the trap length $a_{ho}$ and also half the interaction $g$ then the adiabatic time will be decreased to $(300/16)ms=18.75ms$.

\section{Discussion}
We have shown here, for large interaction strength $|g|$, the original Hamiltonian Eq.~(\ref{originalHamiltonian}) can be map into a spin-chain model governed the by the effective Hamiltonian $H_{\rm eff}$ in the form of Eq.~(\ref{effectiveH}), which is expected to completely describe the physics of the upper branch in the strongly interacting regime. The great advantage of the effective model is that (1) it provides valuable insights into the quantum magnetic properties of strongly interacting one dimensional quantum gases, and (2) it is much easier and more efficient to solve in comparison to the original many-body Hamiltonian. We have benchmarked the static properties of the effective model with several unbiased methods (see Fig.~\ref{Eg1N3} and Fig.~\ref{deltaGgfig}). 

As we have mentioned earlier, recently several other groups have obtained the same spin-chain effective Hamiltonian using a variational method \cite{Volosniev2014_1,Volosniev2014_2,Deuretzbacher2014,Levinsen2014}. Our perturbational approach is inspired by the similar technique used to construct effective spin models from Hubbard Hamiltonian in the large-$U$ limit. Using this technique, the super-exchange interaction arises naturally. The Hubbard Hamiltonian describes lattice systems. Our work thus broadens this approach to a continuum model. From the perturbation calculation presented in this work, we may readily obtain many-body wave functions accurate to order $1/g$. Furthermore, it is in principle possible to extend the perturbation approach to higher orders to obtain more accurate results. These features will be exploited in the future to study more detailed properties of the system.

\begin{figure}[h!]
\includegraphics[width=8.cm]{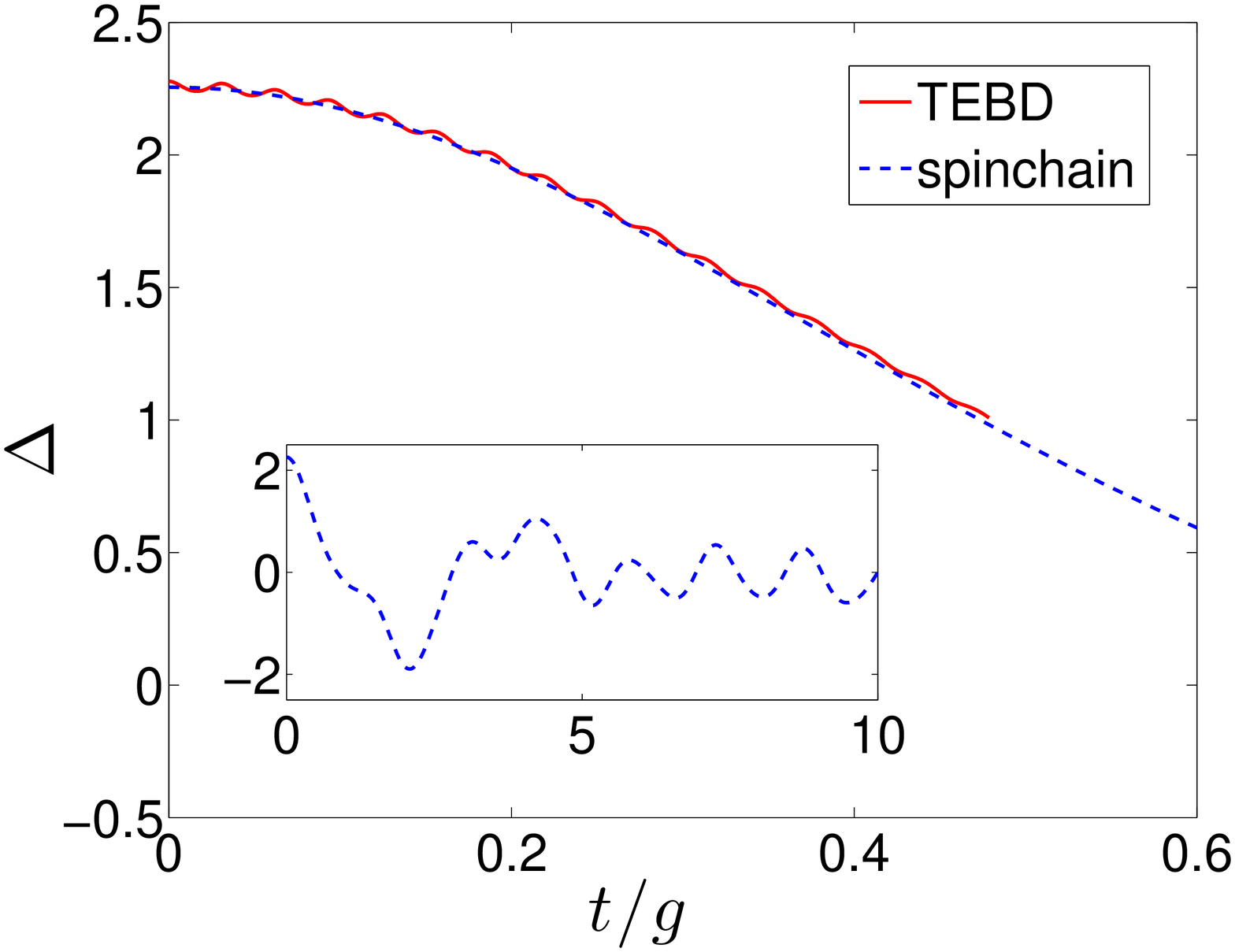}
\caption{(Color online) Quench dynamics for $(N_{\uparrow},N_{\downarrow})=(2,2)$. The initial state is prepared as the ground state with $1/g=0.01$ and $G=0.05$. At $t=0$, $G$ is set to zero, and the system starts to evolve in time. The red solid line is $\Delta(t)$ calculated using the TEBD method, and the blue dashed line is calculated using the effective spin-chain model. The inset figure shows the evolution for much longer time under the spin-chain model. In our trap units, $t/g$ is in units of $\sqrt{m/(\hbar^3 \omega^3)}$.}\label{quenchDynamics}
\end{figure}

In Fig.~\ref{quenchDynamics} we present another example. Here we consider a quench dynamics in which the system is initially prepared in the ground state with $1/g=0.01$ and $G=0.05$. At $t=0$, the spin gradient is suddenly turned off and the evolution of the center-of-mass separation between the two spin species $\Delta$ is calculated by solving the time-dependent Schr\"{o}dinger equation. We solve the Schr\"{o}dinger equaton using both the effective spin-chain model governed by $H_{\rm eff}$, and the TEBD method governed by the original many-body Hamiltonian. As can be seen from Fig.~\ref{quenchDynamics}, the effective model nicely reproduces the TEBD result. We therefore demonstrated that the spin-chain model can be applied to study the dynamics of the system. This example also serves to showcase the advantages of the effective model in the dynamical situation: due to its smaller Hilbert space, it can capture much longer time scale behavior of the system. Furthermore, it takes a few days to obtain the TEBD result as displayed in Fig.~\ref{quenchDynamics}, in comparison to a few tens of seconds for the spin-chain result.

\begin{figure}
\includegraphics[width=8.cm]{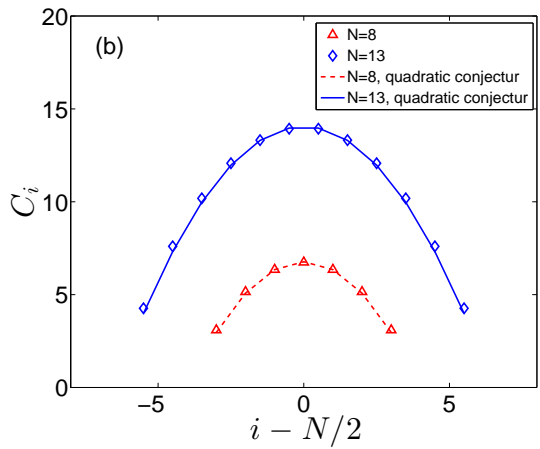}
\caption{(Color online) Dimensionless coefficients $C_i$ for $N=8$ and 13, calculated using the Monte Carlo integral method (Veges Algorithm \cite{Cuba4.0}). The solid lines are obtained using the approximate expression (\ref{conj}).}\label{CiFig1}
\end{figure}

Another great advantage of the spin-chain model is its wide applicability \cite{Volosniev2014_1}. The effective Hamiltonian (\ref{effectiveH}) is valid for spinful fermions, and by changing the minus sign in front of the exchange operator ${\cal E}_{i,i+1}$, it describes strongly interacting bosons. The coefficients $C_i$, as given in Eq.~(\ref{ci}), only depend on the total number of atoms and the external trapping potential, and are independent of whether the particles are bosons or fermions, nor are they dependent on the single particle spin $s$. The formalism to derive the effective Hamiltonian is independent of particle numbers $N$. Hence it works for any $N$. However, for $N$ particles, each coefficient $C_i$ invovles an $N$-dimensional integral, which becomes quite difficult to evaulate as $N$ increases. In Ref.~\cite{Levinsen2014}, the authors conjectured that, for a harmonic trap, these coefficients are given by
\begin{equation}
C_i = K \,\frac{-(i-N/2)^2+N^2/4}{N(N-1)/2} \,,
\label{conj}
\end{equation}
where $K$ is the Tan contact for the AFM state corresponding to $(N_\uparrow, N_\downarrow)=(N-1, 1)$. In Fig.~\ref{CiFig1}, we plot the calculated $C_i$ for $N=8$ and 13 (symbols), in comparison with the above expression (lines), and find good agreement. Hence, at least for harmonic trapped systems, once we know the Tan contact, all the $C_i$ coefficients can be obtained approximately using Eq.~(\ref{conj}). We should also remark that recent experimental progress has made it possible to investigate few-particle cold atom systems with well controlled particle numbers in the lab \cite{few1, few2}.

Finally we comment that we have considered here a system of 1D trapped spinful particles with strong contact interaction, and assumed that the interaction is spin-independent (i.e., SU($2s+1$) symmetric), characterized by a single interaction parameter $g$. It is possible, within the framework of the perturbation method developed here, to generalize the formalism into a situation with spin-dependent interaction strengths, as long as all interaction strengths are sufficiently large \cite{Volosniev2014_2}. Finally, it is also possible to generalize our work to Bose-Fermi mixtures \cite{Girardeau2007,Lelas2009}, which can be compared with recent few-body studies of such mixtures \cite{Garcia-March2014_1,Garcia-March2014_2,Garcia-March2013,Garcia-March2014_3}. We will consider these generalizations in a future work.

\section*{\uppercase{Acknowledgments}}
We thank Xiaoling Cui and Xi-Wen Guan for helpful discussions. This work was supported by NSF and the Welch Foundation (Grant No. C-1669). L.G. acknowledges support from the Tsinghua University Initiative Scientific Research Program.

\appendix
\section{Convention for the permutation operators}
\label{con}

This section contains the convention about the permutation operators and its action on spatial and spin wave functions. A permutation operator $P$ can be expressed as
\begin{align}
\left(\begin{array}{cccc}
1 & 2 & \cdots & N\\
P_{1} & P_{2} & \cdots & P_{N}
\end{array}\right)\,,
\end{align}
which means that the original particle index $i$, after the permutation, is changed into $P_i$.

The action of the permutation operator $P$ on a spatial wave function is defined by
\begin{align}
P\,\psi(x_{1},x_{2}\cdots x_{N})=\psi(x_{P_{1}},x_{P_{2}}\cdots x_{P_{N}})\,.
\end{align}
Similarly its action on a spin wave function is defined by
\begin{align}
P\,\chi(\sigma_{1},\sigma_{2}\cdots\sigma_{N})=\chi(\sigma_{P_{1}},\sigma_{P_{2}}\cdots\sigma_{P_{N}})\,,
\end{align}
where $\sigma_i$ stands for the spin state for $i$th particle. The spin wave function $\chi$ is a rank-$N$ SU($n$) tensor with $n=2s+1$, if all the particles are spin-$s$ particles.

A general spin state can be written as superpositon of basis tensors (or spin Fock states). A basis tensor can be written as $\delta_{\sigma_{1}f_{1}}\delta_{\sigma_{2}f_{2}}\cdots\delta_{\sigma_{N}f_{N}}\equiv\ket{f_{1}f_{2}\cdots f_{N}}$, which means the $i$th spin is in $f_{i}$ state.
By definition, the permutation operator acting on a spin basis yields
\begin{align}
\begin{split}
P\ket{f_{1}f_{2}\cdots f_{N}}
 & = P\delta_{\sigma_{1}f_{1}}\delta_{\sigma_{2}f_{2}}\cdots\delta_{\sigma_{N}f_{N}}\\
 & = \delta_{\sigma_{P_{1}}f_{1}}\delta_{\sigma_{P_{2}}f_{2}}\cdots\delta_{\sigma_{P_{N}}f_{N}}\\
 & = \delta_{\sigma_{1}f_{P_{1}^{-1}}}\delta_{\sigma_{2}f_{P_{2}^{-1}}}\cdots\delta_{\sigma_{N}f_{P_{N}^{-1}}}\\
 & = \ket{f_{P_{1}^{-1}}f_{P_{2}^{-1}}\cdots f_{P_{N}^{-1}}}\,.
\end{split}
\end{align}

We denote ${\cal E}_{i,j}$ as the exchange permutation operator, which simply exchanges indices  $i\leftrightarrow j$:  
\begin{equation} {\cal E}_{i,j}\ket{f_1 \cdots f_i \cdots  f_{j} \cdots f_N}=\ket{f_1 \cdots f_{j} \cdots f_i \cdots f_N} \,.
\end{equation} 
We also denote the symbol $(m\cdots n)$ as a loop permutation operator, which, if $m\geq n$ ($m\leq n)$,  permutes the indices by $m\leftarrow m+1\leftarrow m+2\cdots \leftarrow n\leftarrow m$ ($m\leftarrow m-1\leftarrow m-2\cdots\leftarrow n\leftarrow m$).

\section{Second-order degenerate perturbation theory}
\label{deg}

Consider a Hamiltonian
\begin{align}
H=H_{0}+V\,,
\end{align}
where $H_{0}$ is the unperturbed Hamiltonian, which possess a degenerate manifold with eigen-energy $E^{(0)}$. We define ${\cal P}_0$ the projection operator onto this degenerate subspace. $V$ represents the perturbation Hamiltonian. To calculate the zeroth-order wave function and first-order energy correction, we need to diagnalize ${\cal P}_0V{\cal P}_0$ in ${\cal P}_0$ subspace. Suppose that the first-order energy spectrum still contains a degenerate manifold with energy $E^{(1)}$, and we define ${\cal Q}_0$ as the projection operator onto this remaining degenerate subspace (obviously ${\cal Q}_0\subset {\cal P}_0)$. To lift the degeneracy in ${\cal Q}_0$, we have to consider second-order perturbation.

To distinguish the states in ${\cal Q}_0$ by energy, we need to diagonalize
the following operator in the ${\cal Q}_0$ subspace.
\begin{align}\label{2ndH}
{\cal Q}_0V{\cal P}_1\frac{1}{E^{(0)}-H_{0}}{\cal P}_1V{\cal Q}_0\,,
\end{align}
where ${\cal P}_1=1-{\cal P}_0$ is the complimentary space to ${\cal P}_0$ \cite{2ndOp}. By doing this, we can obtain the zeroth-order wave functions $\ket{l^{(0)}}$ and second-order energy correction $E^{(2)}$. To calculate the first-order wave function correction  $\ket{l^{(1)}}$, we can use
the following two formulas:
\begin{align}\label{P1wf}
{\cal P}_1\ket{l^{(1)}}={\cal P}_1\frac{1}{E^{(0)}-H_{0}}{\cal P}_1V{\cal Q}_0\ket{l^{(0)}}\,,
\end{align}
\begin{align}\label{Q1wf}
{\cal Q}_1\ket{l^{(1)}}={\cal Q}_1\frac{1}{E^{(1)}-V}{\cal Q}_1V{\cal P}_1\frac{1}{E^{(0)}-H_{0}}{\cal P}_1V{\cal Q}_0\ket{l^{(0)}}\,,
\end{align}
where ${\cal Q}_1={\cal P}_0-{\cal Q}_0$ is the complimentary space of ${\cal Q}_0$ in ${\cal P}_0$. The first-order wave function correcton within the ${\cal Q}_0$ subspace can be fixed to be zero, because we have a freedom before normalize the total wave function.

\begin{widetext}
\section{Derivation of the effective spin-chain model}
\label{spin}

For $N$ particles with contact interaction in a trap, the Hamiltonian is given in the main text as Eq.~(\ref{originalHamiltonian}):
\begin{equation}
H=\underbrace{\sum_{i=1}^N\left[-\frac{1}{2}\frac{\partial^2}{\partial x_i^2}+V(x_i)\right]}_{H_f}+\underbrace{g\sum_{i<j}\delta(x_i-x_j)}_{H_{\rm int}}\,.
\end{equation}
In the strongly interacting regime, we take the interaction Hamiltonian $H_{\rm int}$ as the unperturbed Hamiltonian, and the single-particle Hamiltonian $H_f$ as perturbation. The ground state $H_{\rm int}$ is degenerate with energy $E_{\rm int}^{(0)}=0$.

This section details how to derive the second-order perturbation effective Hamiltonian (\ref{2ndH}), with $H_0$ replaced by $H_{\rm int}$ and $V$ replaced by $H_f$, into a spin-chain model. First consider the operator ${\cal P}_1H_f{\cal Q}_0$ acting on an arbitrary state in ${\cal Q}_0$ in the form of Eq.~(\ref{BFwave}) in the main text, which we label here by $\ket{\chi} = \sum_P(-1)^PP(\varphi_{A}\theta^{1}\otimes\chi) $:
\begin{align}
\begin{split}
 {\cal P}_1H_f{\cal Q}_0\ket{\chi}
 &= {\cal P}_1\sum_{i=1}^{N}\left[-\frac{1}{2}\partial_{i}^{2}+V(x_{i})\right] \sum_{P}(-1)^{P}P(\varphi_{A}\theta^{1}\otimes\chi)\\
 &= {\cal P}_1\sum_{P}(-1)^{P}P\left\{ \sum_{i=1}^{N}\left[-\frac{1}{2}\partial_{i}^{2}+V(x_{i})\right] (\varphi_{A}\theta^{1})\otimes\chi)\right\} \\
 &= {\cal P}_1\sum_{P}(-1)^{P}P\left\{ \sum_{i=1}^{N}\frac{1}{2}\left[-2\partial_{i}\varphi_{A}\partial_{i}\theta^{1}-\varphi_{A}\partial_{i}^{2}\theta^{1}\right]\otimes\chi\right\} \\
% &= \sum_{P}(-1)^{P}P\left\{ \sum_{i=1}^{N}\frac{1}{2}\left[-2\partial_{i}\varphi_{A}\partial_{i}\theta^{1}-\partial_{i}(\varphi_{A}\partial_{i}\theta^{1})+\partial_{i}\varphi_{A}\partial_{i}\theta^{1}\right]\otimes\chi\right\} \\
 &=  {\cal P}_1\sum_{P}(-1)^{P}P\left\{ \sum_{i=1}^{N}\frac{1}{2}\left[-\partial_{i}\varphi_{A}\partial_{i}\theta^{1}\right]\otimes\chi\right\} \,.
\end{split}
\end{align}
At the third equal sign we have used the fact that ${\cal P}_1$ projects out
the wave function belonging to the subspace ${\cal Q}_0$. And at the final equal
sign we have used $\varphi_{A}\partial_{i}\theta^{1}=0$, because $\partial_{i}\theta^{1}$
generates $\delta$-functions at $x_i=x_{i\pm 1}$ and the Slater determinant $\varphi_A|_{x_i=x_{i\pm 1}}=0$.

Now let us see how $\partial_{i}\theta^{1}$ generates $\delta$-functions. The sector function $\theta^1$ can be written into a chain product of step functions:
\begin{align}
\theta^1=\theta(x_{2}-x_{1})\theta(x_{3}-x_{2})\cdots\theta(x_{i}-x_{i-1})\theta(x_{i+1}-x_{i}) \cdots\theta(x_{N}-x_{N-1})\,.
\end{align}
We therefore have
\begin{align}
\begin{split}
\partial_{i}\theta^{1} &= \partial_{i}\left[\theta(x_{2}-x_{1})\theta(x_{3}-x_{2})\cdots\theta(x_{i}-x_{i-1})\theta(x_{i+1}-x_{i})\cdots\theta(x_{N}-x_{N-1})\right]\\
 &= \delta(x_{i}-x_{i-1})\left[\theta(x_{2}-x_{1})\theta(x_{3}-x_{2})\cdots\theta(x_{i-1}-x_{i-2})\theta(x_{i+1}-x_{i})\cdots\theta(x_{N}-x_{N-1})\right]\\
 & \;\;\;\; -\delta(x_{i+1}-x_{i})\left[\theta(x_{2}-x_{1})\theta(x_{3}-x_{2})\cdots\theta(x_{i}-x_{i-1}) \theta(x_{i+2}-x_{i+1})\cdots\theta(x_{N}-x_{N-1})\right]\,,
\end{split}
\end{align}
which we rewrite in a simplified notation as
\begin{align}
\partial_{i}\theta^{1}=\delta(x_{i}-x_{i-1})\theta_{[i,i-1]}^{1}-\delta(x_{i+1}-x_{i})\theta_{[i+1,i]}^{1}\,,
\end{align}
where $\theta_{\left[i,i-1\right]}^{1}=\theta^1/\theta(x_i-x_{i-1})$ is the reduced sector function.

Let us now consider the summation $\sum_{i=1}^{N}\frac{1}{2}\left[-\partial_{i}\varphi_{A}\partial_{i}\theta^{1}\right]$.
Since $\partial_{i}\theta^{1}$ and $\partial_{i+1}\theta^{1}$ both
generate $\delta(x_{i}-x_{i+1})$, they can be paired up:
\begin{align}
\begin{split}
    \sum_{i=1}^{N}\frac{1}{2}\left[-\partial_{i}\varphi_{A}\partial_{i}\theta^{1}\right]
 &= \frac{1}{2}\left[-\sum_{i=2}^{N}\partial_{i}\varphi_{A}\delta(x_{i}-x_{i-1})\theta_{[i,i-1]}^{1}+\sum_{i=1}^{N-1}\partial_{i}\varphi_{A}\delta(x_{i+1}-x_{i})\theta_{[i+1,i]}^{1}\right]\\
 &= \frac{1}{2}\sum_{i=1}^{N-1}(\partial_{i}\varphi_{A}-\partial_{i+1}\varphi_{A})\delta(x_{i+1}-x_{i})\theta_{[i+1,i]}^{1}\\
 &= \sum_{i=1}^{N-1}\partial_{i}\varphi_{A}\delta(x_{i+1}-x_{i})\theta_{[i+1,i]}^{1}\,,
 \end{split}
\end{align}
where, in the last step, we have used $\partial_{i+1}\varphi_{A}\delta(x_{i+1}-x_{i})={\cal E}_{i+1,i}\left[\partial_{i+1}\varphi_{A}\delta(x_{i+1}-x_{i})\right]=-\partial_{i}\varphi_{A}\delta(x_{i+1}-x_{i})$,
where ${\cal E}_{i+1,i}$ is an exchange operator that exchanges the index $i+1\leftrightarrow i$. Untill now, we have shown that in the identity spatial sector $x_{1}< x_{2}<\cdots< x_{N}$, the operator ${\cal P}_1H_f{\cal Q}_0$ generates $(N-1)$ $\delta$-functions of neighboring spatial coordinates. The whole expression for ${\cal P}_1H_f{\cal Q}_0\ket{\chi}$ is then
\begin{align}\label{intermState}
{\cal P}_1H_f{\cal Q}_0\ket{\chi}=\sum_{P}(-1)^{P}P\left\{\sum_{i=1}^{N-1}\partial_{i} \varphi_{A}\delta(x_{i+1}-x_{i})\theta_{[i+1,i]}^{1}\otimes\chi\right\} \,.
\end{align}
Next we act ${\cal P}_1(E_{\rm int}^{(0)}-H_{\rm int})^{-1}{\cal P}_1$ on Eq.~(\ref{intermState}). Use the fact that when more than two particles are at a same position, all the $\partial_i\varphi_{A}$'s will vanish and so will Eq.~(\ref{intermState}), we can deal with the $N(N-1)/2$ $\delta$-functions in $H_{\rm int}$ and in Eq.~(\ref{intermState}) separately, which means,
\begin{align}\label{intermState1}
\begin{split}
{\cal P}_1\frac{1}{E_{\rm int}^{(0)}-H_{\rm int}}{\cal P}_1H_f{\cal Q}_0\ket{\chi}
=&-\frac{1}{g}{\cal P}_1\sum_{P}(-1)^{P}P\left\{\sum_{i=1}^{N-1}\partial_{i}\varphi_{A}\frac{\delta(x_{i+1}-x_{i})}{\delta(x_{i+1}-x_{i})}\theta_{[i+1,i]}^{1}\otimes\chi\right\} \\
=&-\frac{1}{g}{\cal P}_1\sum_{P}(-1)^{P}P\left\{\sum_{i=1}^{N-1}\partial_{i}\varphi_{A}\theta_{[i+1,i]}^{1}\otimes\chi\right\}\,.
\end{split}
\end{align}
%Note that there is a pathological term, which is $\delta(x_{i+1}-x_i)/\delta(x_{i+1}-x_i)$. One may question what is the value of it if  $x_{i+1}-x_{i}\ne0$. However, note that there is a projection operator ${\cal P}_1$ that constrains $x_{i+1}-x_{i}=0$. Also it is helpful to imagine the delta function as a infinite height limit of a block function with unit area.

In the final step we act $\bra{\chi'}{\cal Q}_0H_f{\cal P}_1$ on Eq.~(\ref{intermState1}). $\bra{\chi'}{\cal Q}_0H_f{\cal P}_1$ is the hermitian conjugate of a wave function having the form of Eq.~(\ref{intermState}) with a different spin state $\chi'$ but the same $\varphi_A$. Look at Eq.~(\ref{intermState}), since each spatial sector has $N-1$ terms, where each term is composed of a $\delta$-function and a reduced sector function, there will be totally
$N!(N-1)$ terms appearing in this expression. However, only $N!(N-1)/2$
terms are of different $\delta$-function and reduced sector function. For example,
consider a sector $P$ (which labels the sector $x_{P_{1}}<\cdots x_{P_{i}}< x_{P_{i+1}}\cdots< x_{P_{N}}$) and one of its neighbouring sectors $P'=P{\cal E}_{i,i+1}$ (which labels the sector $x_{P_{1}}<\cdots x_{P_{i+1}}< x_{P_{i}}\cdots< x_{P_{N}}$), they both possess the term $\delta(x_{P_{i+1}}-x_{P_{i}})\theta_{\left[P_{i+1},P_{i}\right]}^{P}$. There is also another way to think about this, there are totally $N(N-1)/2$ different $\delta$-functions and for each $\delta$-function there are $\left(N-1\right)!$ different reduced sector functions, so totally $N(N-1)/2\cdot\left(N-1\right)!=N!(N-1)/2$ different terms. Those different terms are orthogonal to each other, because they have different $\delta$-functions and reduced sector functions as well as the fact that when more than two particles are at a same position, $\partial_i\varphi_{A}$ will vanish. For example, one of those $N!(N-1)/2$ terms belonging to sectors $P$ and $P'$ may be
\begin{align}
\begin{split}
 & (-1)^{P}P\left\{ \partial_{i}\varphi_{A}\delta(x_{i+1}-x_{i})\theta_{[i+1,i]}^{1}\otimes\chi'\right\} +(-1)^{P'}P'\left\{ \partial_{i}\varphi_{A}\delta(x_{i+1}-x_{i})\theta_{[i+1,i]}^{1}\otimes\chi'\right\} \\
 &= (-1)^{P}P\left\{ \partial_{i}\varphi_{A}\delta(x_{i+1}-x_{i})\theta_{[i+1,i]}^{1}\otimes\chi'\right\} -(-1)^{P}P {\cal E}_{i,i+1}\left\{ \partial_{i}\varphi_{A}\delta(x_{i+1}-x_{i})\theta_{[i+1,i]}^{1}\otimes\chi'\right\} \\
 &= (-1)^{P}P\left\{ \partial_{i}\varphi_{A}\delta(x_{i+1}-x_{i})\theta_{[i+1,i]}^{1}\otimes\left[1-{\cal E}_{i,i+1}\right]\chi'\right\} \,.
\end{split}
\end{align}
Equation (\ref{intermState1}), similar to $\bra{\chi'}{\cal Q}_0H_f{\cal P}_1$, also has $N!(N-1)/2$ orthogonal terms corresponding to different reduced sector functions and `$\delta$-functions', as the projection operator ${\cal P}_1$ plays the role of the $\delta$-functions. So finally, the matrix elements of the second-order perturbation effective Hamiltonian Eq.~(\ref{2ndH}) can be evaluated as
\begin{align}
\begin{split}
&\braket{\chi'|{\cal Q}_0H_f{\cal P}_1\frac{1}{E_{int}^{(0)}-H_{int}}{\cal P}_1H_f{\cal Q}_0|\chi} \\
=&-\frac{1}{g}\int\prod_{j=1}^Ndx_{j}\left\{\frac{1}{2}\sum_P(-1)^{P}P\left[\sum_{i=1}^{N-1} \partial_{i}\varphi_{A}\delta(x_{i+1}-x_{i})\theta_{[i+1,i]}^{1}\otimes \left[1-{\cal E}_{i,i+1}\right]\chi'\right]\right\}^{\dagger}\\
&\times {\cal P}_1\left\{\frac{1}{2}\sum_{P'}(-1)^{P'}P'\left[\sum_{i=1}^{N-1}\partial_{i}\varphi_{A}\theta_{[i+1,i]}^{1}\otimes \left[1-{\cal E}_{i,i+1}\right]\chi\right]\right\}\\
=&-\frac{1}{g}\int\prod_{j=1}^Ndx_{j}\frac{1}{2}N!\sum_{i=1}^{N-1}\left|\partial_{i}\varphi_{A}\right|^2 \delta(x_{i+1}-x_{i})\theta_{[i+1,i]}^{1}\otimes\left\{\left[1-{\cal E}_{i,i+1}\right]\chi'\right\}^{\dagger} \left\{\left[1-{\cal E}_{i,i+1}\right]\chi\right\}\\
=&-\frac{1}{g}\sum_{i=1}^{N-1}N!\int\prod_{j=1}^Ndx_{j}\left|\partial_{i} \varphi_{A}\right|^2\delta(x_{i+1}-x_{i})\theta_{[i+1,i]}^{1}\otimes \chi'^{\dagger}\left[1-{\cal E}_{i,i+1}\right]\chi \,.
\end{split}
\end{align}
An effective spin-chain model $H_{\rm eff}$ is therefore obtained
\begin{align}\label{spinchain}
H_{\rm eff}=-\frac{1}{g}\sum_{i=1}^{N-1}C_{i}\left[1-{\cal E}_{i,i+1}\right] \,,
\end{align}
where
\begin{align}\label{Ci}
C_{i}=N!\int\prod_{j}dx_{j}\left|\partial_{i}\varphi_{A}\right|^2\delta(x_{i+1}-x_{i})\theta_{[i+1,i]}^{1}\,.
\end{align}
The above derivation is valid for fermions. In the case of bosons, a general many-body wave function can be written as
\begin{equation}
\Psi(x_{1}\cdots x_{N},\sigma_{1}\cdots\sigma_{N})=\sum_{P}P\left[\varphi_{A}\theta^1\otimes\chi\right] \,.
\end{equation}
Following the same procedure as above, we end up with an effective Hamiltonian as
\begin{align}
H_{\rm eff}=-\frac{1}{g}\sum_{i=1}^{N-1}C_{i}\left[1+{\cal E}_{i,i+1}\right] \,,
\end{align}

\section{One-body density matrix}
\label{den}
Given a many-body wave function $\Psi$, the one-body density matrix is defined as:
\begin{align}
\begin{split}
\rho_{\sigma'\sigma}(x',x)
&=\sum_{\sigma_2\cdots\sigma_N}\int dx_2\cdots dx_N \,\Psi^*(x',x_2\cdots x_N,\sigma',\sigma_2\cdots\sigma_N)\Psi(x,x_2\ldots x_N,\sigma,\sigma_2\cdots\sigma_N)\,.
%\\
%&=\frac{1}{N}\braket{\Psi|c_{\sigma'}^{\dagger}(x')c_{\sigma}(x)|\Psi}
\end{split}
\end{align}
For fermionic systems whose wave function takes the form of Eq.~(\ref{BFwave}) in the main text,
\begin{align}
\Psi=\sum_{P}(-1)^{P}P\left[\varphi_{A}\theta^{1}\otimes\chi\right]=\varphi_{A}\sum_{P} \left[\theta^{P}\otimes\chi\right]\,,
\end{align}
where $\theta^{P}=P\theta$ is the sector function (generalized step function) for the sector labled by permutation operator $P$, the one-body density matrix can be written as
\begin{align}
\rho_{\sigma'\sigma}(x',x)=\sum_{\sigma_2\cdots\sigma_N}\int dx_2\cdots dx_{N}\varphi_A'^*\varphi_A\sum_{P'P}\theta'^{P'}\theta^{P}\otimes(P'\chi'^*)(P\chi)\,,
\end{align}
where $\varphi_A'^*=\varphi_A^*(x',x_{2}\cdots x_{N})$, $\varphi_A=\varphi_A(x,x_{2}\cdots x_{N})$, $\theta'^{P'}=P'\theta(x',x_{2}\cdots x_{N})$, $\theta^P=P\theta(x,x_{2}\cdots x_{N})$, $\chi'^*=\chi^*(\sigma',\sigma_{2}\cdots\sigma_{N})$, and $\chi=\chi(\sigma,\sigma_{2}\cdots\sigma_{N})$.
A permutation $P$ can be written as $P_{2-N}\cdot(1\cdots m)$, where $(1\cdots m)$ is the loop permutation operator defined in Appendix~\ref{con}, and $P_{2-N}$ is a permutation operator acting on indices $2,3\cdots N$. This means first move particle $1$ to position $m$ by a loop permutation, and then permute the remaining $N-1$ particles. Similarly, $P'$ can be written as $P'=P'_{2-N}\cdot(1\cdots n)$. The summation over $P'$ and $P$ can then be written into another form:
\begin{align}
\begin{split}
\rho_{\sigma'\sigma}(x',x)
&=\sum_{\sigma_2\cdots\sigma_N}\int dx_2\cdots dx_N\varphi_{A}'^*\varphi_A\sum_{m,n}\sum_{\substack{P'_{2-N},\\P_{2-N}}}\theta'^{P'_{2-N}\cdot(1\cdots m)}\theta^{P_{2-N}\cdot(1\cdots n)}\otimes\left[P'_{2-N}\cdot(1\cdots m)\chi'^*\right]\left[P_{2-N}\cdot(1\cdots n)\chi\right]\\
&=\sum_{\sigma_2\cdots\sigma_N}\int dx_2\cdots dx_{N}\varphi_{A}'^*\varphi_{A}\sum_{mn}\sum_{P_{2-N}}\theta'^{P_{2-N}\cdot(1\cdots m)}\theta^{P_{2-N}\cdot(1\cdots n)}\otimes\left[P_{2-N}\cdot(1\cdots m)\chi'^*\right]\left[P_{2-N}\cdot(1\cdots n)\chi\right]\\
&=\sum_{mn}(N-1)!\int dx_{2}\cdots dx_{N}\varphi_{A}'^*\varphi_{A}\theta'^{(1\cdots m)}\theta^{(1\cdots n)}\otimes\sum_{\sigma_{2}\cdots\sigma_{N}}\left[(1\cdots m)\chi'^*\right]\left[(1\cdots n)\chi\right]\,.
\end{split}
\end{align}
The second equal sign follows the fact that if $P'_{2-N}\ne P_{2-N}$, $\theta^{'P'_{2-N}}(x_{2}\cdots x_{m},x'\cdots x_{N})\theta^{P_{2-N}}(x_{2}\cdots x_{n},x\cdots x_{N})=0$, and the third equal sign uses the fact that $\sum_{\sigma_{2}\cdots\sigma_{N}}\int dx_{2}\cdots dx_{N}$ is invariant under $P_{2-N}$. So the one-body density matrix can be separated into a spatial part and a spin part
\begin{align}
\rho_{\sigma'\sigma}(x',x)=\sum_{m,n}\rho_{m,n}(x',x)S_{m,n}(\sigma',\sigma)\,,
\end{align}
where the spatial part
\begin{align}
\rho_{m,n}(x',x)=(N-1)!\int dx_{2}\cdots dx_{N}\,\varphi_{A}'^{*}\varphi_{A}\,\theta'^{(1\cdots m)}\theta^{(1\cdots n)}\,,
\end{align}
is simply the one-body density matrix of a system of spinless fermions for the spatial sector $x_{2}<x_{3}\cdots x_{m}<x'\cdots x_{n}<x\cdots x_{N}$ ($m<n$, for example). And the spin part
\begin{align}
S_{m,n}(\sigma',\sigma)=\sum_{\sigma_{2}\cdots\sigma_{N}}\left[(1\cdots m)\chi^{*}\right]\left[(1\cdots n)\chi\right]=\braket{\chi|c_{m}^{\dagger}(\sigma')(m\cdots n)c_{n}(\sigma)|\chi}\,,
\end{align}
where $(m\cdots n)$ is the loop permutation operator, and $c_{m}^{\dagger}(\sigma)$ can be regarded as fermion (or hard core boson) creation operators, which is just a formal symbol to select out the spin states. For bosons, simply change the sectored spinless fermionic one-body density matrix to bosonic one by $\rho_{m,n}^B=(-1)^{m-n}\rho_{m,n}^F$.

\section{Green's function results for $(N_{\uparrow},N_{\downarrow})=(1,1)$}
\label{2b}
The Hamiltonian of two particles in a one dimensional harmonic trap with a spin dependent magnetic gradient is
\begin{align}\label{N2Hamiltonian}
H
=-\frac{1}{2}\frac{\partial^2}{\partial x_1^2}-\frac{1}{2}\frac{\partial^2}{\partial x_1^2}+\frac{1}{2}x_2^2+\frac{1}{2}x_2^2+g\delta(x_1-x_2)-Gx_1\sigma_1^z-Gx_2\sigma_2^z\,.
\end{align}
In the absence of the magnetic gradient (i.e., $G=0$), there exists an exact solution to the problem \cite{Busch1998}. Here we generalize this solution in the presence of the magnetic gradient. To this end, we make a transformation of operators by making spatial and spin coordinates operators into Jacobi coordinates:
\begin{align}\label{cmrel}
X_1=\frac{x_1-x_2}{\sqrt{2}},\;X_2=\frac{x_1+x_2}{\sqrt{2}},\;S_1=\frac{\sigma_1^z-\sigma_2^z}{\sqrt{2}}, \;S_2=\frac{\sigma_1^z+\sigma_2^z}{\sqrt{2}}\,.
\end{align}
The transformation rules of other operators such as $\partial/\partial x$ can be obtained from them. The Hamiltonian can be separated into the center-of-mass motion part and the relative motion part:
\begin{align}\label{N2HamiltonianSep}
H=-\frac{1}{2}\frac{\partial^2}{\partial X_2^2}+\frac{1}{2}X_2^2-GS_2X_2-\frac{1}{2}\frac{\partial^2}{\partial X_1^2}+\frac{1}{2}X_1^2-GS_1X_1+\frac{g}{\sqrt{2}}\,\delta(X_1)\,.
\end{align}
For center-of-mass motion, it is a simple harmonic oscillator with center shifted by $GS_2$. For relative motion, it is a simple harmonic oscillator with center shifted by $GS_1$ plus a $\delta$-function potential at the origin. We can first let particle 1 to be spin up and particle 2 to be spin down, then anti-symmetrize the wave function in the end. In this case, $S_2=0$ and $S_1=\sqrt{2}$ are fixed. The eigen wave functions for the center-of-mass motion are still simple harmonic oscillator eigen-functions. What matters is the relative motion part. After a coordinate shift $X_1-GS_1\rightarrow X_1$, The relative motion Hamiltonian can be written as
\begin{align}\label{relHN2}
H_{\rm rel}&=-\frac{1}{2}\frac{\partial^2}{\partial X_1^2}+\frac{1}{2}X_1^2+\frac{g}{\sqrt{2}}\,\delta(X_1+GS_1)\,,
\end{align}
which includes a simple harmonic oscillator part and a $\delta$-function source term. For this relative Hamiltonian, use the one-body Green's function
\begin{align}
G(E;X_1,X'_1)=\sum_{i=0}^{\infty} \frac{1}{E-E_i}\phi_i(X_1)\phi_i^*(X'_1)\,,
\end{align}
where $E_i=i+1/2$ and $\phi_i$ are the single particle harmonic oscillator eigen-energies and eigen-wave functions, respectively. The corresponding Lippmann-Schwinger equation for the relative wavefunction is given by
\begin{align}\label{wfGreen}
\begin{split}
\varphi(X_1)&=\int dX_1G(E;X_1,X'_1)\frac{g}{\sqrt{2}}\,\delta(X_1+GS_1)\varphi(X'_1)\\
&=\frac{g}{\sqrt{2}}\,G(E;X_1,-GS_1)\varphi(-GS_1)\,.
\end{split}
\end{align}
We just got the expression for the relative wave function, where $\varphi(-GS_1)$ is a constant can be determined by normalization of $\varphi(X_1)$. And the relative energy must satisfy
\begin{align}\label{energyEq}
G(E;-GS_1,-GS_1)=\frac{\sqrt{2}}{g}\,.
\end{align}
Note that, when $G=0$, the Green's function method fails at $E=E_i$, and for $E\ne E_i$ the left hand side of Eq.~(\ref{energyEq}) has an analytical form \cite{Busch1998}. Actually the solution of fully symmetric spin wave function (necessarily assoticated with fully anti-symmetric spatial wave function) which has $E=E_i$ should be complemented to the Green's function solution. However, for $G\neq 0$, there is no such pathological behavior for the Green's function method. Also to be noted is that for one $E$, there could be only one $1/g$ for which Eq.~(\ref{energyEq}) is satisfied. This means, for relative motion, there could be only one bound state. However, this is no longer true for three particles, because the Lippmann-Schwinger equation for three particles is an integral equation and there can exist infinitely many bound states for three particles' relative motion.

Finally, substitute back $X_1\rightarrow X_1-GS_1$, after anti-symmetrization, the total wavefunction for two fermions is given by
\begin{align}\label{totalwavefunction}
\frac{1}{\sqrt{2}}\Psi_{cm}(X_2)\left[\varphi(X_1-G\sqrt{2}) \ket{\uparrow\downarrow}-\varphi(-X_1-G\sqrt{2})\ket{\downarrow\uparrow}\right]\,.
\end{align}
The center-of-mass separation between the two spins, $\Delta=\braket{x_1\sigma_1^z+x_2\sigma_2^z}/2$, can be calculated as
\begin{align}\label{N2Delta}
\Delta=\frac{1}{\sqrt{2}}\int dX_{1}X_{1}\left|\varphi(X_{1})\right|^{2}+G\,,
\end{align}
where $\varphi(X_1)$ is decided by Eq.~(\ref{wfGreen}), which dependents on $G$ and $E$, where $E$ is dependent on $G$ and $1/g$ by Eq.~(\ref{energyEq}). The first term in Eq.~(\ref{N2Delta}) is from interplay between the interaction and the magnetic gradient, while the second term in Eq.~(\ref{N2Delta}) is due to the harmonic trap shift induced by the magnetic gradient.

\end{widetext}


\begin{thebibliography}{99}
\bibitem{2ndOp}  This form is a little different from the conventional second-order degenerate perturbation effective Hamiltonian where ${\cal P}_1$ should be $1-{\cal Q}_0$. But note that ${\cal Q}_0$ is obtained by first diagonalizing $V$ in ${\cal P}_0$, so $({\cal P}_0-{\cal Q}_0)V{\cal Q}_0={\cal Q}_1V{\cal Q}_0=0$.
\end{thebibliography}
\end{document}